\documentstyle[aps,prb,psfig,epsf]{revtex} \newcommand{\kdp}{\mbox{KH$_2$PO$_4$} }
\begin{document}

\twocolumn[\hsize\textwidth\columnwidth\hsize\csname@twocolumnfalse\endcsname

\title{First-principles study of ferroelectricity and isotope effects in H-bonded KDP crystals}

\author{S. Koval$^{\rm (1)}$, J. Kohanoff,$^{\rm (2)}$,
        J. Lasave$^{\rm (1)}$, G. Colizzi$^{\rm (2)}$,
        and R. Migoni $^{\rm (1)}$ }

\address{
         $^{1)}$ Instituto de F\'{\i}sica Rosario, Universidad
                 Nacional de Rosario,\\
                 27 de Febrero 210 Bis, 2000 Rosario, Argentina. \\
         $^{2)}$ Atomistic Simulation Group, The Queen's University,
                 Belfast BT7 1NN, Northern Ireland.\\
        }
\date{\today}
\maketitle

\begin{abstract}

By means of extensive first-principles calculations we studied the
ferroelectric phase transition and the associated isotope effect
in KH$_2$PO$_4$ (KDP). Our calculations revealed that the
spontaneous polarization of the ferroelectric phase is due to
electronic charge redistributions and ionic displacements which
are consequence of proton ordering, and not vice versa. The
experimentally observed double-peaked proton distribution in the
paraelectric phase cannot be explained by a dynamics of only
protons. This requires, instead, collective displacements within
clusters that include also the heavier ions. These tunneling
clusters can explain the recent evidence of tunneling obtained
from Compton scattering measurements. The sole effect of mass
change upon deuteration is not sufficient to explain the huge
isotope effect. Instead, we find that structural modifications
deeply connected with the chemistry of the H-bonds produce a
feedback effect on tunneling that strongly enhances the
phenomenon. The resulting influence of the geometric changes on
the isotope effect agrees with experimental data from neutron
scattering. Calculations under pressure allowed us to analyze the
issue of universality in the disappearance of ferroelectricity
upon compression. Compressing DKDP so that the distance between
the two peaks in the deuteron distribution is the same as for
protons in KDP, corresponds to a modification of the underlying
double-well potential, which becomes 23 meV shallower. This energy
difference is what is required to modify the O-O distance in such
a way as to have the same distribution for protons and deuterons.
At the high pressures required experimentally, the above feedback
mechanism is crucial to explain the magnitude of the geometrical
effect.

\end{abstract}

\vspace{0.5cm}
]

\begin{small}\begin{normalsize}\end{normalsize}\end{small}\section{Introduction.}

Potassium dihydrogen phosphate (KH$_2$PO$_4$, or KDP) crystals are
a key component in quantum electronics. They are widely used in
controlling and modulating the frequency of laser radiation in
optoelectronic devices, amongst other uses such as TV screens,
electrooptic deflector prisms, i nterdigital electrodes, light
deflectors, and adjustable light filters.
 Besides the obvious technological interest, KDP is also interesting from
 a fundamental point of view.  KDP is a prototype ferroelectric (FE) crystal
  belonging to the family
of hydrogen-bonded ferroelectrics, which was extensively studied
in the past.~\cite{blizec,lingla} Their PO$_4$ molecular units are
linked by hydrogen bonds, and ferroelectricity appears to be
connected to the behavior
 of the protons in these H-bonds. Normal (H$_2$O) ice is the most prominent
 member of this family,~\cite{Ben98,Bra99} which also includes other compounds
  like PbHPO$_4$,~\cite{Loc85} and squaric acid (C$_4$H$_2$O$_4$).~\cite{samara}
  What makes KDP particularly interesting is the possibility of growing quite large,
 high-quality single crystals from solution, thus making it very suitable
 for experimental studies. Indeed, a large wealth of experimental data has
 been accumulated during the second half of the past century.~\cite{blizec,lingla}

Phosphates in KDP are linked through approximately planar H-bonds
forming a three-dimensional network. In the paraelectric (PE)
phase at high temperature, the H-atoms occupy with equal
probability two symmetrical positions along the H-bond separated a
distance $\delta$, which characterizes the so-called disordered
phase. Below the critical temperature T$_c$ $\approx$ 122 $K$, the
protons localize into one of the symmetric sites, thus leading to
the ordered FE phase. Here, the spontaneous polarization P$_s$
appears perpendicular to the proton ordering plane, and the PO$_4$
tetrahedra distort. The proton configuration in this phase is
depicted in Fig. 1; each PO$_4$ unit has two covalently bonded and
two H-bonded hydrogen atoms, following Slater's ice rules
\cite{slater}. The oxygen atoms that bind covalently to the
hydrogen are called {\it acceptor} (O2 in Fig. 1), and those
H-bonded are called donors (O1 in Fig. 1).

A striking feature common to all H-bonded ferroelectrics is
undoubtedly the huge isotope effect observed upon deuteration. In
fact, the deuterated compound (DKDP) exhibits a T$_c$ about two
times larger than KDP. This giant effect was first explained by
the quantum tunneling model proposed in the early
sixties~\cite{Bli60}. Within the assumption of interacting,
single-proton double wells, this model proposes that individual
protons tunnel between the two wells. Protons are more delocalized
than deuterons, thus favoring the onset of the disordered PE phase
at a lower T$_c$. Improvements of the above model include coupling
between the proton and the K-PO$_4$
dynamics.~\cite{Kob68,Mat82,Koj88,Bus98} These models have been
validated {\it a posteriori} on the basis of their predictions,
although there is no direct experimental evidence of tunneling.
Only very recent neutron Compton scattering experiments seem to
indicate the presence of tunneling \cite{Rei02}. However, the
connection between tunneling and isotope effect remains unclear,
in spite of recent careful experiments.~\cite{Ike94}

On the contrary, a series of experiments carried out since the
late eighties ~\cite{Ich87,Tun88,Nel88,McM90,Sel99} provided
increasing experimental evidence that the geometrical modification
of the hydrogen bonds and the lattice parameters upon deuteration
(Ubbelohde effect~\cite{Ubb39}) is intimately connected to the
mechanism of the phase transition. The distance $\delta$ between
the two collective equilibrium positions of the protons (see Fig.
1) was shown to be remarkably correlated with T$_c$.~\cite{McM90}
Actually, it seems that proton and host cage are connected in a
non-trivial way, and are not separable.~\cite{Kru90} These
findings stimulated new theoretical work where virtually the same
phenomenology could be explained without invoking
tunneling.~\cite{Sug91,Sug96,Mer_Rav} However, these theories were
developed at a rather phenomenological level. Only very recently,
the first {\it ab initio} calculations, based on Density
Functional Theory (DFT), were conducted in these
systems.~\cite{comp_mat_sci,ferro02,Zha02,prlkov_02,Liu03} These
approaches have the advantage of allowing for a confident and
parameter-free analysis of the microscopic changes affecting the
different phases in this system.

In this work we investigate, using DFT electronic structure
calculations within the generalized gradient approximation to
exchange and correlation, the relationship between proton
ordering, internal geometry, polarization, tunneling and isotope
effects in KDP (details of the methods used are exposed in Section
II). To this end, consideration of the following questions
naturally arises: (1) What is the microscopic mechanism which
gives rise to the FE instability?, (2) How do local instabilities
lead to the double-site distribution in the PE phase?, (3) What is
the quantum origin of the geometrical effect?, (4) What is the
main cause of the giant isotope effect: tunneling or the
geometrical modification of the H-bonds?, (5) How does pressure
affect the energetics and the structural parameters in the system?

With the aim of shedding light on the above formulated questions,
and on the general problematic in KDP, we conducted different
computational experiments and made a revision of previously
obtained results. First, we carried out electronic structure
calculations in the tetragonal unpolarized phase (PE), forcing
protons to be in the middle of the O-H-O bonds. Calculations in
the polarized phase (FE), with the H ordered off-center, were
performed in both, the tetragonal fixed cell and in the completely
relaxed cell, which is orthorhombic. We studied the structures and
the charge reorganization leading to the FE instability. These
results are presented in Section III. In Section IV, we analyze
global instabilities in KDP to understand the relation between
proton ordering and polarization. To address the tunneling issue,
we studied local instabilities by determining the dependence of
the system energetics upon the proton position in the H-bonds
under various conditions: allowing or not K and P ions
relaxations, and considering also individual proton and small
cluster displacements. Besides, we show in this section a
calculation of the momentum distribution of the proton along the
H-bond in different phases and compare the results with recent
experimental data. Section V is devoted to a thorough study of
quantum fluctuations, and the controversial problem of the isotope
effects. We show how a self-consistent quantum modeling, based on
our first principles calculations, is able to explain the striking
mass dependence of the geometrical effect. In Section VI, we
present calculations of the energetics and the structural
parameters as a function of pressure. We show that the results of
related experiments under pressure are explained by the non-
linear relationship between deuteration and geometric effects,
derived in the previous section. Finally, in Section VII we
discuss the above issues, and elaborate our conclusions.

\section{AB INITIO METHODS}
We have performed ab initio calculations of KDP, within the
framework of DFT, \cite{Hoh64,Koh65} using two different
pseudopotential codes, one based on localized basis sets (LB), and
another using plane waves (PW).

\vspace{-0.3 truecm}
\begin{figure}[b]
\epsfxsize=7.5cm
\epsfysize=4.8cm
\hspace*{0.cm} \epsfbox{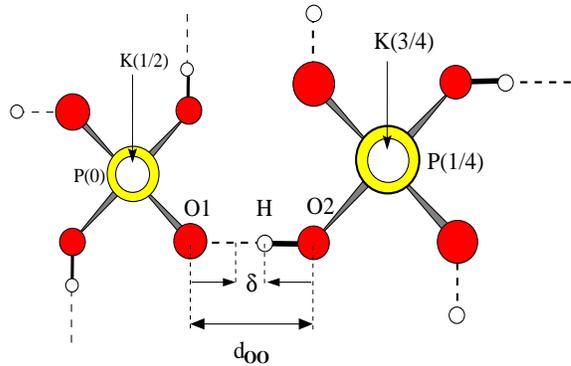}
\vskip 0.cm
\caption{Schematic view of the internal structure of KDP along the tetragonal
axis. The fractional coordinates of P and K atoms along the c axis,
are indicated in brackets. Covalent and H-bonded hydrogens are
connected to corresponding oxygens by full and broken lines, respectively.}
\end{figure}
The LB calculations were carried out using the SIESTA program.
~\cite{Ord96,San97} This is a fully self-consistent DFT method
that employs a linear combination of pseudoatomic orbitals (LCAO)
of the Sankey-Niklewsky type as basis functions~\cite{Sankey}.
These basis functions are strictly confined in real space, what is
achieved by imposing in the pseudoatomic problem (i.e. the atomic
problem where the Coulomb potential has been replaced by the same
pseudopotential that will be used in the solid state), the
boundary condition that the orbitals vanish at a finite cutoff
radius, rather than at infinity as for the free atom. Therefore,
these solutions are slightly different from the free atom case,
and have somewhat larger associated energies because of the
confining potential. The relevant parameter for this approximation
is precisely the orbital confinement energy $E_c$, given by the
energy difference between the eigenvalues of the confined and the
free orbitals. In our calculations, we used a value of
$E_c=50$meV. By decreasing this value further, we checked that we
obtain total energies and geometries with sufficient accuracy. For
the representation of the valence 

\twocolumn[\hsize\textwidth\columnwidth\hsize\csname@twocolumnfalse\endcsname

\begin{table}
\caption{Comparison of the ab-initio (LB and PW) internal structure parameters
with DKDP experimental data~\cite{Nel87} for the different cases considered
in the text. The notation is the same used in the experimental works referred.
$\gamma$ is the relative z-displacement of the K and P atoms from the
equidistant situation (see definition in section IV.A). Distances in~\AA~and
angles in degrees.}
\begin{center}
\begin{tabular}{l|ccc|cc|ccc}
           &\multicolumn{5}{c|}{Tetragonal}&\multicolumn{3}{c}{Orthorhombic}\\
\cline{2-6}
 & \multicolumn{3}{c|}{Unpolarized (UT)} & \multicolumn{2}{c|}{Polarized (PT)}
&  &  &\\
\cline{2-9}
 Structural &  LB   &  PW   &  Exp. &  LB   &  PW   &  LB   &   PW   &  Exp. \\
 parameters &       &       &(234 K)&       &       &       &        &(219 K)\\
\hline\hline
d(P-O2)     & 1.594 & 1.565 & 1.543 & 1.624 & 1.599 & 1.625 & 1.593  & 1.578 \\
d(P-O1)     & 1.594 & 1.565 & 1.543 & 1.571 & 1.536 & 1.569 & 1.528  & 1.509 \\
d$_{OO}$    & 2.422 & 2.418 & 2.522 & 2.465 & 2.497 & 2.480 & 2.491  & 2.533 \\
$\delta$    &   0   &   0   & 0.443 & 0.275 & 0.371 & 0.310 & 0.381  & 0.472 \\
$\gamma$    & 0 &  0    &   0   & 0.072 & 0.107 & 0.082 & 0.120  & 0.130 \\
$<$ O2-P-O2 & 111.2 & 110.6 & 110.5 & 106.4 & 106.2 & 105.6 & 106.3  & 105.7 \\
$<$ O1-P-O1 & 111.2 & 110.6 & 110.5 & 114.5 & 115.3 & 115.1 & 115.8  & 115.7 \\
$<$O1$\cdots$H-O2&177.4&178.3 & 177.1 & 177.0 & 178.9 & 177.3 & 178.9  & 179.8 \\
$\theta$    & 60.2  & 59.4  &  61.6 &  61.7 &  61.6 &  62.8 &  62.0  &  62.3 \\
E$_{gap}$ (eV)& 5.64& 5.78  &       &  5.55 &  5.65 & 5.52  &  5.65  &       \\
\end{tabular}
\end{center}
\end{table}
\vspace{-0.5truecm}
]

\noindent  electrons in the LB method we
used double-zeta bases with polarization functions (DZP). This
means two sets
of orbitals for the angular momenta occupied in the isolated atom,
and one set of orbitals for the first non-occupied angular
momentum (polarization orbitals). Again, this size of the basis
set turns out to be accurate enough for our purposes.

The exchange-correlation energy terms were computed using the
Perdew-Burke-Ernzerhof (PBE) form of the generalized gradient
approximation \cite{Per96}. This type of functionals has already
been used to describe hydrogen-bonded systems, with quite good
accuracy.~\cite{parrinello:water} We also tried the BLYP
functional \cite{blyp}, which gives very good results for
molecular systems. However, the results in the solid state were of
quality inferior than PBE. We used non-local, norm-conserving
Troullier-Martins pseudopotentials \cite{Tro91} to eliminate the
core electrons from the description. We also included nonlinear
core corrections (NLCC) for a proper description of the K ion, due
to an important overlap of the core charge with the valence charge
density in this atom. We checked this approximation by comparing
with a K$^+$ pseudopotential that includes the 3$s$ and 3$p$
shells explicitly in the valence, as semicore states (9 explicit
electrons per K atom). The semicore results are very closely
reproduced by the core-corrected calculations. For the real-space
grid used to compute numerically the Coulomb and
exchange-correlation integrals~\cite{Ord96,San97}, we used an
equivalent energy cutoff of 125 Ry.

The pseudopotential PW calculations were carried out with the
PWSCF code ~\cite{Bar01}, with the same exchange-correlation
functional and pseudopotentials, except for the core-corrected K,
where we used a semicore pseudopotential for K$^+$ (9 electrons
per K atom). The plane wave expansion was cut off at a maximum PW
kinetic energy of 150 Ry. Such a high cutoff was necessary to
obtain convergence in the internal degrees of freedom,
particularly the hydrogen-bonded units. The PWSCF code also allows
for the computation, within the linear response regime, of
vibrational and dielectric properties such as phonon frequencies,
Born effective charges and dielectric constant. Phonon
eigenvectors were used to calculate the total energy curves under
pressure, by constraining the optimization to motions preserving
the pattern of the ferroelectric normal mode, which is related to
the parameter $\delta$.

The PE phase of KDP has a body-centered tetragonal ($bct$)
structure with 2 formula units per lattice site (16 atoms). For
the LB calculations that describe homogeneous distortions, we used
the conventional $bct$ cell (4 formula units), but doubled along
the tetragonal $c$ axis. This supercell comprises 8 formula units
(64 atoms). A larger supercell is required to describe local
distortions. To this end, we used the equivalent conventional
$fct$ cell (containing 8 formula units, and axes rotated through
45 degrees with respect to the conventional $bct$ cell), also
doubled along the $c$-axis (128 atoms). The LB calculations were
conducted using a $\Gamma$-point sampling of the Brillouin zone
(BZ). This choice of sampling proved sufficient provided the large
supercells used in the calculations.

Most of the calculations have been done using the LB approach
which, within the approximations described above, turned out into
quite a fast computational procedure, compared to PW calculations.
As a test, we checked the LB approach against the PW results. The
PW calculations were carried out on the 16-atom $bct$ unit cell,
with a BZ sampling consisting of 8 centered Monkhorst-Pack
k-points. This number of points was checked for convergence, and
proved sufficient. The results for the geometrical parameters are
reported in table I. It can be seen that the LB values are of
quality comparable to the PW results. The differences can be
attributed mainly to the approximation made with the confinement
of the orbitals in the LB calculations.

\section{CHARACTERIZATION OF THE STRUCTURES AND CHARGE FLOW MECHANISMS}

We first optimized the structure with paraelectric phase symmetry.
To this end, we constrained the H-atoms to remain centered in the
O-H-O bonds, and fixed the lattice parameters to the experimental
values of the deuterated compound (DKDP) at T$_c$+5~K ({\it a = b
=} 7.459 {\AA} and {\it c =} 6.957 \AA) in the conventional $bct$
cell.~\cite{Nel87} The choice of DKDP instead of KDP for the
comparison with experiment is based on the fact that nuclear
quantum effects, which are neglected in the first-principles
calculations, are less important. Optimization of all the atomic
positions leads to what we call the {\it unpolarized tetragonal}
(UT) structure. This can be interpreted as an average of the true
paraelectric phase. In fact, according to experimental data, in
this latter the H-atoms are observed with equal probability in two
symmetric off-centered positions along the H-bonds. In Table I we
compare the relevant structural parameters resulting from both
types of calculations and also experimental data. The agreement
between the two theoretical approaches is quite good -- thus
validating the later use of the LB approach --, and their
comparison with experiment is very satisfactory, except for the
O-O distance $d_{OO}$ which, specially in the UT case, turns out
to be too short. ~\cite{comp_mat_sci} This delicate issue will be
discussed below.

Maintaining the lattice parameters and constraining the K and P
atoms to their centered positions in the UT structure, we next
allowed for H off-center relaxation towards the ordered
configuration sketched in Fig. 1. The O-O distance is also
optimized. In this way we obtained a H off-center shift $\delta/2
= 0.154$~{\AA} and an O-O distance of 2.472~\AA. We will show
below in more detail that the H off-centering produces an
electronic charge redistribution from the neighborhood of the O2
atoms towards that of the O1 atoms. As a consequence, unbalanced
forces are generated that favo pairing of the K and P atoms along
the $z$ axis, on the charge-excess side (O1) of the PO $_4$ units.
The former observation indicates that, constraining the K and P to
their centered positions, does not prevent the H atoms from
abandoning the center of the H-bonds. The centered position for
the H atoms is always unstable, as we will show in the next
Section.

The next step was to relax also the positions of the K and P
atoms, thus leading to the {\it polarized tetragonal} (PT)
structure, whose geometrical parameters are listed in Table I.
This PT structure is not yet the ground state, because the
ferroelectric distortion is coupled to a shear strain mode. It is
this acoustic mode that becomes soft before the FE mode, thus
piloting a structural transition to an orthorhombic phase, which
is very similar to the PT. This coupling can be observed in the
$\sigma_{xy}$, off-diagonal components of the calculated stress
tensor. According to this observation, we relaxed again all the
internal degrees of freedom, but now fixing the simulation cell to
the experimental orthorhombic structure of DKDP at $T_c-10~K$.
This corresponds to lattice parameters $a=10.598$~\AA,
$b=10.496$~\AA~ and $c=6.961$~{\AA} in the conventional $fct$
cell. ~\cite{Nel87} The calculated geometrical parameters, which
are close to those of the PT structure, are shown in the last
three columns in Table I compared to experimental data.

In the experimental orthorhombic structure, however, the stress
tensor is diagonal but not isotropic. This indicates that, if the
lattice parameters were also to be optimized, the $b/a$ ratio
would be different from the experimental value. In addition, the
isotropic part of the stress (the pressure) is non-zero, thus
indicating a small difference in equilibrium volume between
calculations and experiment. In general the agreement is quite
reasonable, again with the exception of the O-O distance, which is
0.1 \AA~too small in the UT structure. This is a very important
issue, because the potential for the deuterons (or protons) in the
H-bond is extremely sensitive to the O-O distance~\cite{scheiner}.
In the present DFT-PBE calculations for the UT structure this
distance is 2.42 {\AA}, i.e. 0.1 {\AA} shorter than the
experimental value~\cite{Nel87}. This difference, which can even
change the shape of the potential felt by the deuteron in the
H-bond, cannot be fully attributed to the optimization for
centered deuterons. In fact, it persists when we optimize the
structure in the orthorhombic FE phase, although slightly reduced
(2.49 {\AA} vs. 2.53 {\AA}). One possible reason are quantum
nuclear effects. Our calculations are for clamped nuclei,
corresponding to infinite deuteron mass. If quantum dynamics of
deuterons was to be included, it would slightly increment this
discrepancy because nuclear delocalization favors shorter H-bonds.
This can also be seen from the fact that the experimental O-O
distance for protons is shorter than for deuterons (Ubbelohde
effect). Therefore, the inclusion of quantum effects would imply
even shorter O-O distances. It is neither a problem of the
pseudopotential approach, which has been tested against
all-electron calculations.

We conclude, then, that the main origin of the underestimation of
the O-O bond length is in the approximate character of the
exchange-correlation functional. In fact, calculations for related
gas-phase systems like H$_3$O$_2^-$ indicate a similar 0.06 {\AA}
underestimation when comparing GGA values to correlated quantum
chemical calculations~\cite{grimm}. Moreover, present test
calculations for the water dimer also indicate and underestimation
of the d$_{OO}$ distance by 0.06 {\AA} with respect to
experimental values~\cite{dyke}.

Therefore, the differences in the H-bond geometry app-

\twocolumn[\hsize\textwidth\columnwidth\hsize\csname@twocolumnfalse\endcsname

\begin{table}
\caption{\small Changes q(PT) - q(UT) in the Mulliken orbital and bond
overlap populations in going from the UT to the PT configuration, in units
of e/1000.}
\begin{center}
\begin{tabular}{ccccc|cccc}
 O1 &  O2 &  P &  K &  H  &  O1$\cdots$H & O1-P &  O2-H & O2-P\\
\hline
+82 & -58 & -8 & -3 & -17 &    -91     &  46    &  70  & -44 \\
\end{tabular}
\end{center}
\end{table}

\vspace{-0.5truecm}

\begin{figure}
\centerline{
\psfig{figure={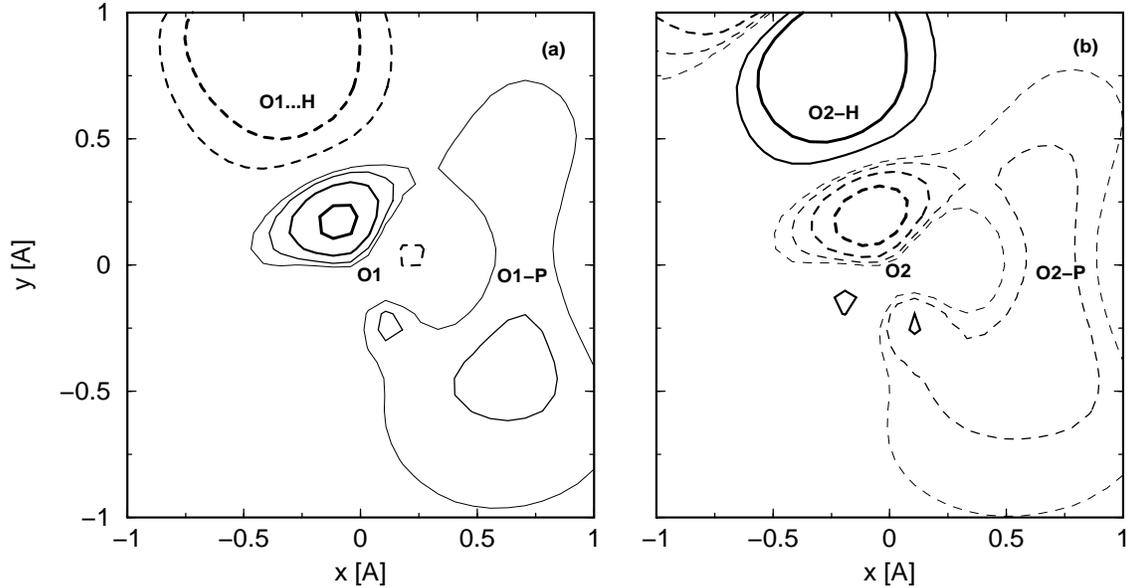},angle=270,width=15cm,height=8cm}}
\caption{Differential charge density contours $\Delta \rho({\bf r})$
in the planes containing the following atoms: (a) P-O1$\cdots$H,
(b) P-O2-H. Labels O1 and O2 denote the positions of the respective
nuclei, positioned at (0,0). Labels O2-P and O1-P indicate the position
of the center of the corresponding bonds. The same convention is used
for the O2-H  and O1$\cdots$H bonds. Positive (negative) contours are in
solid (dashed) lines.
The thickest lines represent an absolute value of $2.96 \times 10^{-3}$
e\AA$^{-3}$. The thinner lines are obtained by successively halving this
value, down to $3.70 \times 10^{-4}$ e\AA$^{-3}$.}
\end{figure}
]

\noindent ear to be mainly due to
the approximate character of the exchange-correlation term. Unfortunately, at
present, a sufficiently well validated and efficient scheme to go beyond GGA
is lacking. Therefore, in order to avoid problems derived from this feature
in the study of the stability of local cluster distortions in next Section,
we decided to fix the O-O distances in the host to the experimental values.

With the purpose of analyzing the charge redistributions produced
by the ordered proton off-centering, we computed the changes in
the Mulliken orbital and bond-overlap populations in going from
the UT to the PT configuration, as shown on Table II. Mulliken
populations depend strongly on the choice of the basis set.
Differences, however, are much less sensitive.

An increase of the charge localized around O1 can be clearly
observed; the main contribution ($\approx 70$\%) is provided by a
decrease in the O2 charge.~\cite{comp_mat_sci} The trends observed
in Table II are confirmed by the charge density difference $\Delta
\rho({\bf r})=\rho_{PT} ({\bf r})-\rho_{UT}({\bf r})$. In Fig. 2a
and 2b we plot cuts of the above quantity in the planes determined
by the atoms P-O1$\cdots$H and P-O2-H , respectively. A combined
analysis of both, Table II and Fig. 2a, indicates a significant
enhancement of the population of the O1 atom, accompanied by a
smaller increment in the O1-P orbitals. This happens at the
expenses of the population of the O1$\cdots$H and O2-P overlap
orbitals, and the population of the O2 atom. Therefore, as two
H-atoms move away from O1 and other two approach O2, the O1-H bond
weakens and the O2-H bond strengthens. The charge localizes mostly
around O1 and, to a lesser extent, in the P-O1 orbitals. This is
consistent with the increase of $d(O1-O2)$ and the decrease of
$d(P-O1)$ reported in Table I. The contrary occurs in the vicinity
of the O2 atom, as indicated by the orbital and bond-overlap
populations in Table II and the contours in Fig. 2b. The overall
effect is a flow of electronic charge from the O2 side of the
tetrahedron towards the O1 side, and a concomitant modification of
its internal geometry. The charge redistribution is rather local,
and gives rise to a polarization composed by electronic and ionic
contributions~\cite{popova}. This polarization, whose origin can
therefore be traced back to the off-centering of the H-atoms in
the perpendicular plane, is intimately linked to ferroelectricity.
In fact, the combined motion of all the atoms and the concomitant
electronic redistribution corresponds to an unstable phonon in the
UT structure~\cite{giuseppe:phd}. When this phonon mode freezes
into one of the two stable minima, we obtain the PT structure
which has a polarization, and is thus ferroelectric. A schematic
view of this combined effect is presented in Fig. 3.
\begin{figure}
\centerline{
\psfig{figure={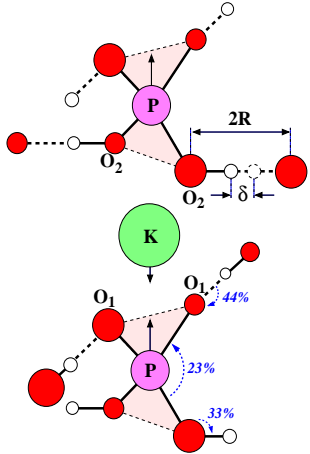},width=4.cm,height=6.cm}}
\vspace*{0.4 truecm}
\caption{Schematic view, perpendicular to the $c$-axis, of the atomic
motions (solid arrows) and electronic charge redistributions
(dotted curved arrows), happening upon off-centering of the H-atoms.
The percentages of the total charge redistributed
are also shown for the charge-transfers occurring
between different orbitals and atoms.}
\label{schema}
\end{figure}

A further confirmation of these ideas comes from the saturated
polarization which was obtained from linear response calculations
of Born effective charges. \cite{giuseppe:phd,Bar01} Considering
the eigenvector of the FE mode $e_{\alpha} (i)$, where $\alpha$
indicates the Cartesian coordinate and $i$ the ion, we calculated
the dynamical effective charge in the z direction:

\begin{equation}
Z^*_{z}(FE)=  \sum_i \sum_{\alpha} Z^*_{z \alpha}(i)
~\frac{e_{\alpha} (i)} {\sqrt{m_i}} = 1.6~e. \label{efcharge}
\end{equation}

\noindent The effective charge components in the x and y
directions vanish. $Z^*_z(FE)$ is multiplied by the FE-mode
amplitude corresponding to the stable minimum, giving rise to a
saturation polarization of $P_s=3.25 \mu C/cm^2$, slightly lower
than experimental values. Analyzing the individual contributions,
we observe that a substantial part arises, in fact, from the P
ions. However, this contribution is only a 40\% of the total
polarization. The other 60\% arises from the H atoms through a
non-diagonal $xz$ (or $yz$) component of the effective charge
tensor. This component is $Z^*_{xz}(H)=0.6~e$ while
$Z^*_{zz}(P)=3~e$, but the displacement of the H atoms in the $x$
(or $y$) axis is more than five times larger than that of the P
ions along the $z$ axis. This, and the fact that there are twice
as many H than P in the unit cell, explains why H contributes as
much as P to the polarization. The interesting observation is that
the H atoms move in the plane perpendicular to the $z$ axis.
Therefore, their off-diagonal contribution can only be due to
electronic polarization effects. Although the effective charges of
the K and O atoms are not null, their small displacements lead to
negligible contributions to the spontaneous polarization.

\section{GLOBAL AND LOCAL FERROELECTRIC INSTABILITIES}

\subsection{Correlation Between Proton Ordering and Polarization}

In the previous section we showed that there is an instability of
the system towards the PT structure as hydrogens collectively move
away from the centers of the O-H-O bond.  From our simulations, we
observe that when the protons are constrained to remain centered
in the H-bonds, the K and P atoms are stable in their centered
positions. However, centering the heavy atoms does not imply the
centering of the H-atoms. Protons, in fact, are never stable at
their centered positions. This provides a strong evidence that the
origin of ferroelectricity is in the off-center ordering of the
protons, and that proton off-centering and ferroelectricity are
very correlated phenomena.

To identify the driving mechanism of the ferroelectric
instability, we analyzed the relationship between proton ordering
and polarization. To this purpose, we investigated the {\it ab
initio} potential energy surface (PES) as a function of the proton
off-centering parameter $\delta=d_{OO}-2d_{OH}$, and the K-P
relative displacement along the $c$-axis, which we quantify in
terms of the parameter $\gamma=d_{PP}-2d_{KP}$, with $d_{KP}$ the
smallest K-P distance. It is worth mentioning here that $\gamma$
is a measure of polarization, since a test calculation provided us
a linear relationship between these quantities.

We fully relax the oxygen positions for each chosen
$(\delta,\gamma)$ pair, and plot the energy contours of the
bidimensional PES in the inset to Fig. 4. The characteristics of
this PES are as follows: it exhibits a saddle point at $\delta =
\gamma = 0$, and two equivalent minima at $(\delta, \gamma) \simeq
\pm (0.3, 0.15) \AA$. On the one hand, from the energy contours it
can be seen that at $\delta = 0$ (centered protons) there is no
instability for any value of $\gamma$, i.e. the crystal is stable
against polarization ($\gamma \neq 0$) unless the protons are
ordered off-center. On the other hand, even for vanishing $\gamma$
(polarization) the energy minimum corresponds to a finite
$\delta$, i.e. protons are always collectively unstable at the
H-bond center. This is further visualized in Fig. 4, where we plot
the energy profiles as a function of $\delta$ for different fixed
values of $\gamma$. For $\gamma=0$, the energy profile exhibits a
double-well in the $\delta$ coordinate with a barrier of $\simeq
6$ meV per molecular unit. For increasing values of $\gamma$ the
minima are always at $\delta\neq 0$, up to a value of $\gamma
\approx 0.02$ {\AA}, where one of the two minima completely
disappears. Therefore, we conclude from the above considerations
that the source of the ferroelectric instability is the H
off-centering, and not viceversa.

\begin{figure}[tb]
\vspace*{-1.4 truecm}
\centerline{
\psfig{figure={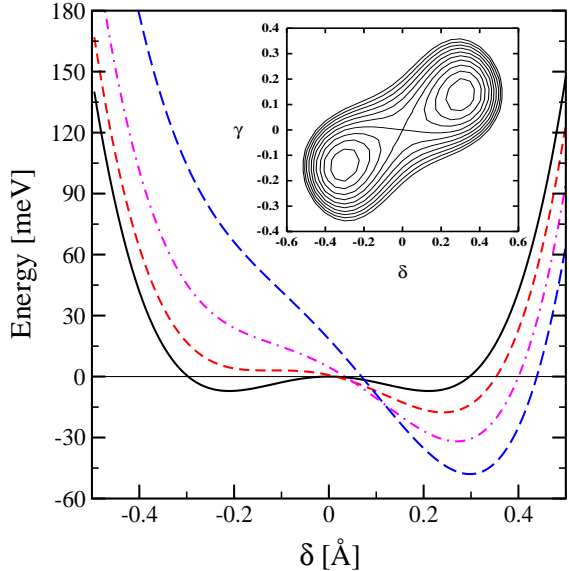},angle=0,width=9.cm,height=12.cm}}
\vspace{-3. truecm}
\caption{Energy profiles as a function of $\delta$, for values of
$\gamma=$ 0 (solid line), 0.02 (short-dashed), 0.05 (dot-dashed) and
0.1 (long-dashed) {\AA}. The inset shows equispaced energy
contours (step = 13.6 meV/ \kdp unit). The minima at
$(\delta, \gamma) \simeq \pm (0.3, 0.15) \AA$
lie $\simeq 50$ meV below the saddle point at (0,0)}
\label{delgam}
\end{figure}

\subsection{Local Instabilities, Quantization in Small Clusters, and
the Nature of the Paraelectric Phase}

We now address the microscopic origin of the observed proton
double-occupancy in the PE phase \cite{Nel87}, which is an
indication of the order-disorder character of the transition. This
phenomenon can be ascribed either to static or thermally activated
dynamic disorder, or to tunneling between the two sites. Any of
these possibilities requires the search for instabilities with
respect to correlated but localized H motions in the PE phase,
including also the possibility of heavy-ions relaxation. In fact,
correlated motions of a large number of protons become
increasingly unlikely in a tunneling scenario, because this
implies higher barriers and heavier effective masses, thus
reducing the tunneling probability. To analyze localized
distortions we consider increasingly larger clusters embedded in a
host paraelectric matrix. For the reasons exposed in the previous
Section, the host is modeled by protons centered between oxygens,
and the experimental structural parameters (including the O-O
distances) of KDP at T$^{KDP}_c$+5~K (127 K).~\cite{Nel87} In
order to assess the effect of the volume increase observed upon
deuteration, we also analyze the analogous case of D in DKDP by
expanding the host structural parameters to the corresponding
experimental values at T$^{DKDP}_c$+5~K (234 K). ~\cite{Nel87}

We analyze results for different clusters comprising N hydrogens
(deuteriums): (a) N=1 H(D) atom, (b) N=4 H(D) atoms which connect
a PO$_4$ group to the host, (c) N=7 H(D) atoms localized around
two PO$_4$ groups, and (d) N=10 H(D) atoms localized around three
PO$_4$ groups. For all these clusters we consider correlated
motions with the pattern shown in Fig. 1, which are the most
favorable for exhibiting FE instabilities, as it was illustrated
in the previous section. This correlated pattern, is represented
by a single collective coordinate $x$ whose value coincides with
the H(D) off-center displacement $\delta/2$. Two cases are
considered: (i) first, we allow for the motion of H atoms alone,
maintaining all other atoms fixed, (ii) second, we also allow for
the relaxation of the heavy ions K and P, which follow the
ferroelectric mode pattern~\cite{Nel_rev_87,giuseppe:phd}, as
expected. Subsequent quantization of the cluster motion in the
corresponding effective potential allows for the determination of
the importance of tunneling in the disordered phase. Rigorously,
the size dependence should be studied for larger clusters than
those mentioned here. However, it will be shown below that short-
range quantum fluctuations in the PE phase are sufficiently
revealing, especially far away from the critical point.

In Fig. 5 we show, for the clusters considered, the total energy
variation as a function of $x$. For the case of H motions alone,
we do not observe any instability for N=1 and N=4, both in KDP and
DKDP. A small barrier of $\sim 6$ meV appears in DKDP for the N=7
move, as shown in Fig. 5(b) (open squares). This barrier grows up
to $\approx$ 25 meV for the N=10 cluster in DKDP (open circles).

\begin{figure}
\vspace*{-0.5 truecm}
\centerline{
\psfig{figure={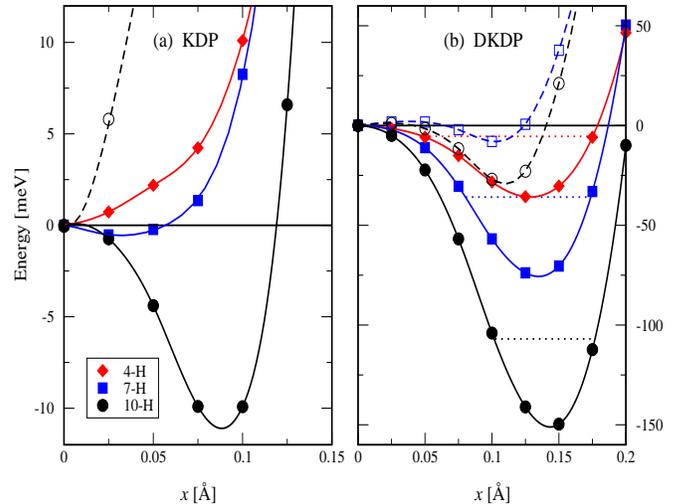},angle=270,width=9.cm,height=8.cm}}
\caption{Energy profiles for correlated local distortions in (a) KDP
and (b) DKDP. Reported are clusters of: 4 H(D) (diamonds), 7 H(D)
(squares), and 10 H(D) (circles). Empty symbols and dashed lines
indicate that only the H(D) atoms move. Motions that involve also
heavy atoms (P and K) are represented by filled symbols and solid
lines. Lines are guide to the eye only.}
\label{local}
\end{figure}

\noindent  However, quantum mechanical calculations of the cluster
levels, which will be described bellow, yield ground states (GS)
quantized above the barriers and, consequently, the absence of
tunneling in those cases (see Fig. 5). In KDP, even the largest
cluster considered is very stable, as indicated by the open
circles in Fig. 5(a). The result for KDP suggests to rule out this
type of motions in the paraelectric phase, because they are
incompatible with double site occupancy.

In a second step we considered also the motion of the heavy atoms
in the above correlated local motions. The situation changed
drastically, as shown by the solid lines and full symbols in Fig.
5(a) and (b). In fact, clusters involving two or more PO$_4$ units
-- cases (c) and (d) above -- exhibit instabilities in both KDP
and DKDP, with a significant barrier in DKDP for case (d), of the
order of $\approx$ 150 meV. We note here that the instability
appears in clusters which are sufficiently large, thus providing a
measure of the FE correlation length. Moreover, the instabilities
are much stronger (and the correlation length accordingly shorter)
in the expanded DKDP lattice, than in KDP.

We treat these clusters quantum-mechanically, by solving the
Schr\"odinger equation for the collective coordinate $x$. This is
done for each cluster in the corresponding effective potentials of
Fig. 5. The effective mass for the local collective motion of the
cluster is calculated as $\mu = \sum_i m_i a_i^2$, where $i$ runs
over the displacing atoms and $m_i$ are their corresponding atomic
masses. $a_i$ is the $i$-atom displacement at the minimum from
their positions in the PE phase, relative to the H(D)
displacement. The effective masses per H(D) calculated for these
correlated motions in different clusters are about $\mu_H \approx
2.3$ ($\mu_D \approx 3.0$) proton masses (m$_p$) in KDP (DKDP),
respectively. The calculation of the GS energy in the {\em heavy
clusters}, leads now to quantized levels below the barriers, as
shown by dotted lines in Fig. 5(b), for all clusters in DKDP. This
is a clear sign for tunneling arising from correlated D motions
involving also heavy ions. These collective motions can be
understood as a local distortion reminiscent of the global FE
mode.~\cite{giuseppe:phd} In KDP, however, even the largest
cluster considered (N=10), has the GS level quantized above the
barrier. The onset of tunneling at a critical cluster size,
provides a rough indication of the correlation volume: it
comprises more than 10 hydrogens in KDP, but no more than 4
deuteriums in DKDP. ~\cite{note} We clearly observe, then, that
the dynamics of the order-disorder transition would involve fairly
large H(D)-clusters together with heavy-atom (P and K)
displacements. Thus, the observed proton double-occupancy is
explained in our calculation by the tunneling of large and {\em
heavy} clusters. The last conclusion is confirmed by the
double-site distribution determined experimentally for the P
atoms.~\cite{McM90e,McM91}

A possible scenario for the FE phase transition would then be the
following: the PE phase is made of clusters of different size,
some of them (the large ones) preserve the FE structure, i.e. do
not exhibit double site H(D) occupancy because the barrier is too
high. Other clusters (smaller) have lower barriers and smaller
effective masses, and thus can tunnel, giving rise to double
occupancy. The effect of temperature is that of modifying the
preferential cluster size, which grows as a measure of the
correlation length on approaching the transition. When the average
cluster size reaches a value in which neither tunneling nor
thermal hopping are anymore allowed, then the phase transition
takes place. Of course, this is a mean field vision, but we
believe that the picture is quite plausible.

\subsection{Momentum Distributions}

Since Blinc's model proposal,~\cite{Bli60} it has been subject of
controversy whether the protons are actually tunneling between the
two equivalent sites along the bridges, or they are localized in
one of these sites and jump to the other through phonon assisted
tunneling in the paraelectric phase. Reiter et al. ~\cite{Rei02}
have recently attempted to elucidate this question by performing
neutron Compton scattering experiments. Due to the much shorter
time scale of this experiment, compared to typical times for
phonon assisted jumps in the paraelectric phase, it is claimed
that it is possible to distinguish between a proton coherently
distributed between the two equivalent sites, and one which is
alternatively occupying one site or the other. In this experiment,
the momentum distribution along the bridge, $n(p)$, has been
obtained by inverting the measured scattering function under
plausible conditions.~\cite{Rei02} Very significant changes in
$n(p)$ are observed when going through the transition, which were
not to be expected if the proton was localized only in one of the
equivalent sites, in both phases. As shown by the solid lines in
Fig. 6, $n(p)$ is considerably narrower in the high temperature
phase, indicating an increase in the spread of the region where
the proton is coherently distributed (the wave packet). More
conclusively, the high temperature distribution shows a zero and a
subsequent oscillation which correspond precisely to a double
peaked spatial wavefunction, i.e. the proton coherently
distributed over both sites along the bond. In contrast, well
below T$_c$, $n(p)$ shows a single and broader maximum at $p=0$,
thus indicating single-site occupancy.

Our calculations for the coherent motion of hydrogen clusters with
fixed heavy ions, in a host of a mean paraelectric phase, indicate
that only very large clusters would exhibit double well potentials
with energy barriers high enough to allow for collective
tunneling. On the other hand, considerably smaller clusters are
able to tunnel if also the heavy ions are allowed to move
coherently with H in KDP, or with D in DKDP. Therefore, Compton
scattering results can be explained if the observed coherent
double peaked distribution of a proton along a bridge is
interpreted as part of a coherent motion together with heavy ions
in a cluster.

Since the largest cluster we treated in KDP is not able to tunnel,
a direct comparation of our momentum distribution calculations in
the PE phase with experiment is not feasible. Instead, we can make
a prediction of what would the $n(p)$ distribution be like, in the
PE phase of DKDP. For this purpose, we considered the
corresponding double-well potentials as functions of the position
of D along the bridge, for the 4-D and 7-D clusters in DKDP. We
calculated the wavefunctions with the cluster effective masses
$\mu =$ 10.4 m$_p$ and 21.4 m$_p$, respectively. The resulting
momentum distributions of DKDP in the PE phase are shown in Fig. 6
(right panel), together with the the experimental curve of KDP for
illustrative purposes. The second oscillation arises from the
quantum coherence of the real space distribution. As the effective
mass of the cluster increases, the second oscillation has larger
amplitudes, while the main oscillation remains unchanged.

In the ferroelectric phase of KDP the proton distribution is
single peaked, corresponding to a single-well anharmonic potential
for each proton. The momentum distribution calculated for a single
proton wave function in such potential is shown by the dashed line
in the left panel of Fig. 6. In order to understand the difference
with the experimental data, we performed another calculation where
the surrounding ions are allowed to relax. This leads to a
shallower potential, but the effective mass also increases. The
result, shown as a dotted line, deviates even more from the data.
The deviation from experiment, observed for the uncorrelated
proton distribution (dashed line), may be due to the broadening
effect of temperature on the distribution of the host ions. It is
worth mentioning that differences between results from the present
calculations and previous preliminary ones~\cite{ferro02} are due
to refinements of the potentials performed presently.

\section{QUANTUM DELOCALIZATION AND THE GEOMETRICAL EFFECT}

\subsection{Geometrical Effect vs. Tunneling}

We now address the origin of the huge isotope effect on T$_c$
observed in KDP and also in the isomorphic H-bonded crystals in
the family. For forty years, starting from the pioneering work of
Blinc, ~\cite{Bli60} the central issue in KDP has been whether
tunneling is or not at the root of the large isotope effect, a
fact that was never rigorously confirmed. Moreover, a crucial set
of experiments pointing against the tunneling picture was recently
conducted by Nelmes and co-workers: by applying pressure, they
conveniently tuned the D-shift parameter $\delta_{DKDP}$ in DKDP
to make it coincide with the H-shift parameter $\delta_{KDP}$ in
KDP, and they observed that T$_c^{DKDP}$ almost coincided with
T$_c^{KDP}$, in spite of the mass difference between D and H in
both systems.~\cite{Nel88,McM90,Nel91} This suggests that the
modification of the H-bond geometry by deuteration -- the {\em
geometrical effect} -- is a central mechanism in the transition,
and is intimately connected with the isotope  effect.

\begin{figure}
\centerline{
\psfig{figure={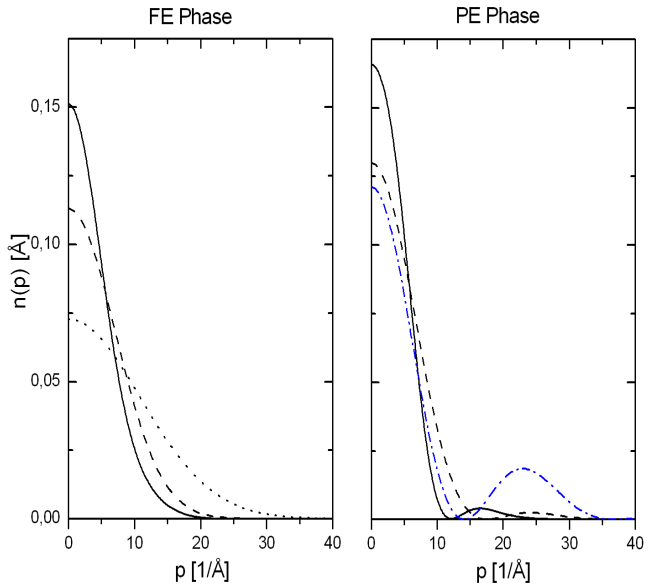},angle=0,width=10.cm,height=8.5cm}}
\vspace{-0.25truecm}
\caption{Momentum distributions along the H-bond in both phases of KDP. Experimental data
are shown in solid lines.~\cite{Rei02}
Left panel (FE phase): calculations for single
uncorrelated H motion (dashed line), and for H correlated with host relaxation (dotted line).
Right panel (PE phase): calculations for 4-D and 7-D cluster dynamics, as explained in text
(dashed and dot-dashed lines, respectively).
}
\label{mom}
\end{figure}

As the cluster size grows (N $\rightarrow \infty$), the tunnel
splitting $\Omega$ vanishes. However, only large clusters are
expected to be relevant for the nearly second-order FE transition
in these systems.~\cite{Sam73} Thus, for large tunneling clusters
the potential barrier is sufficiently large, and the GS levels are
deep enough (see Fig. 5) that the relation $ \hbar \Omega_{H(D)}
\ll K_B T_c$ is satisfied, so much for D as for H. Therefore,
according to the tunneling model, the above relation implies that
a simple change of mass upon deuteration at fixed potential cannot
explain the near duplication of T$_c$. In fact, if we consider the
largest cluster (N=10) in Fig. 5 for DKDP, which is larger than
the crossover length in this system, we have that the GS level for
the deuterated case (calculated with an effective mass of
$10\mu_D=35.4$ m$_p$) is around $E_{GS}=$ -107 meV, well below the
central barrier, and the tunnel splitting amounts to $ \hbar
\Omega_D = 0.34$ K. Modifying the mass at fixed potential
($10\mu_H=25.3$ m$_p$) leads to a tunnel splitting of $ \hbar
\Omega_D = 1.74$ K. Since T$_c^{DKDP} \approx 229 K$, the relation
$\hbar \Omega_{H(D)} \ll K_B T_c$ clearly holds, and the change in
$\Omega$ at fixed potential accounts only for a small change in
T$_c$. This is in agreement with the high-pressure experiments
mentioned above,~\cite{Nel88,McM90,Nel91} where at fixed
structural conditions, the isotope effect in T$_c$ appears to be
rather modest.

Also the geometric effect in the H-bond is very small at fixed
potential. In fact, the proton and deuteron wave functions (WF) in
the DKDP potential for the N=7 cluster, which are reported in Fig.
7 (a), are only slightly different. As a matter of fact, the
distance between peaks as a function of the effective mass at
fixed potential remains almost unchanged, as can be seen by the
square symbols in Fig. 7 (c).

In contrast, the proton WF for the N=7 cluster in the KDP
potential exhibits a single broad peak, as shown in Fig. 7 (a).
But, how can we then explain such a large geometric change in
going from DKDP to KDP? The first observation arises from what is
apparent in Figure 5: energy barriers in DKDP are much larger than
those in KDP, implying that quantum effects are significantly
reduced in the expanded DKDP lattice. In fact, the proton WF in
the DKDP potential has more weight in the middle of the H-bond ($x
\approx 0$) than the deuteron WF (Fig. 7 (a)). That is, due to
zero-point motion, protons are more favored in the H-centered
position than deuterons. This affects the covalency of the bond,
which becomes stronger as the proton moves to the H-bond center,
as discussed in Section III. The geometric change of the O-H-O
bridge, produced by the combined effect of quantum delocalization
and gain in covalency, affects in turn the crystal cohesion. Thus,
the increased probability of the proton to be midway between
oxygens strengthens the O-H-O covalent grip and pulls the oxygen
atoms together, causing a small contraction of the lattice. As
will be shown in the following Section, this contraction has the
effect of decreasing the barrier height, thus making the proton
even more delocalized. This triggers a further contraction of the
lattice, and so on in a self-consistent way. This self-consistent
procedure is finally identified as the phenomenon that makes the
lattice shrink from the larger classical value to the smaller
value found for KDP. This phenomenon, triggered by tunneling and
quantum delocalization, leads to an enhancement of the geometrical
effect. The overall self-consistent effect is eventually much
larger than the deuteration effect obtained at fixed potential,
i.e. at fixed lattice constant.

To estimate an upper limit to that effect, we compared the lattice
parameters and the bridge lengths by carrying out electronic
calculations with classical nuclei (clamped nuclei). This was done
for two different situations: one with the hydrogens forced to
stay in the middle of the H-bond, and the other with the hydrogens
fully off-centered in the FE state of KDP. In the latest case, the
distance between oxygens is $d_{OO} \approx 2.50$ \AA, falling to
$d_{OO} \approx 2.42$ \AA~ when H is centered. In addition, the
lattice volume is contracted by about 2.3 \%. Thus, the proton
centering acts as a very strong attraction center, pulling the two
oxygens together. We estimate that, at the equilibrium volume, the
proton centering creates an equivalent pressure of $\approx$ 20
Kbar. In the true high-temperature PE phase, though, the protons
are not centered in the middle of the H-bonds, but they are
equally distributed on both sides of the bond, thus reducing the
magnitude of the effect.

\subsection{The Isotope Effect: a Nonlinear Self-consistent Phenomenon}

In the previous subsection we discussed how a self-consistent
mechanism combining quantum delocalization, the modification of
the covalency in the bond, and the effect on the lattice
parameters, can account for the large geometric effect observed
upon deuteration. This mechanism is now capable of explaining, at
least qualitatively, the increase in the order parameter and T$_c$
with deuteration. This self-consistent mechanism has obviously its
origin in the difference in tunneling induced by different masses,
but is largely amplified through the geometric modification of
bond lengths and energy scales.

To demonstrate the effect of isotopic substitution via this
self-consistent non- linear mechanism, we constructed the
following simple model: we considered the Schr\"odinger equation
for the clusters with a WF-dependent term added to the bare
potential. The effective potential reads:

\begin{equation}
V_{\rm eff}(x)=V_0(x) - k |\Psi(x)|^2, \label{efpot}
\end{equation}

\noindent where $x$ is the collective coordinate of the cluster
and $V_0(x)$ is a quartic double-well similar to those of Fig. 5.
The term in $|\Psi(x)|^2$ serves as a non-linear feedback in the
model: when the particle is more delocalized, it has more weight
in the middle of the H-bond. Then, $|\Psi(x)|^2$ increases at the
center, the effective barrier is lowered, the particle further
delocalizes, and so on, self-consistently. The bare potential can
be written as

\begin{equation}
V_{0}(x)=E_b^0 \left[ -2 \left( \frac{2x}{\delta^0_{min}}
\right)^2 +
                     \left( \frac{2x}{\delta^0_{min}} \right)^4 \right] ,
\label{barepot}
\end{equation}

\noindent in terms of its energy barrier $E_b^0$ and minima
separation $ \delta^0_{min} $. The parameters values $k$ = 20.2
meV.\AA, $E_b^0$ = 35 meV and $\delta^0_{min}$ = 0.24 {\AA} were
chosen so as to qualitatively reproduce the WF profiles in the
cases of KDP (broad single peak) and DKDP (double peak), for the
same cluster size. Once these parameters are fixed, the WF
self-consistent solutions depend only on the effective mass.
Figure 7(b) shows the WF corresponding to $\mu_D$ (solid line) and
$\mu_H$ (dashed line), which are similar to those calculated from
the ab initio potentials for the N=7 cluster (Fig. 7(a)).

In Fig. 7(c), we show the distance between peaks $\delta_p$ in the
WF as a function of the cluster effective mass $\mu$. Starting
from the finite value for $\mu_D$ (DKDP), $\delta_p$ decreases
remarkably towards lower $\mu$ values, until it vanishes near
$\mu_H$ (KDP) (see circles in Fig. 7(c)). This strong dependence
of $\delta_p$ on the mass is in striking contrast with the very
weak dependence obtained at fixed DKDP potential and geometry
(square symbols). Such a large mass dependence, can now explain
the large isotope effect found in KDP, via an amplified and
self-consistent geometrical modification of the H-bond.

\begin{figure}
\centerline{
\psfig{figure={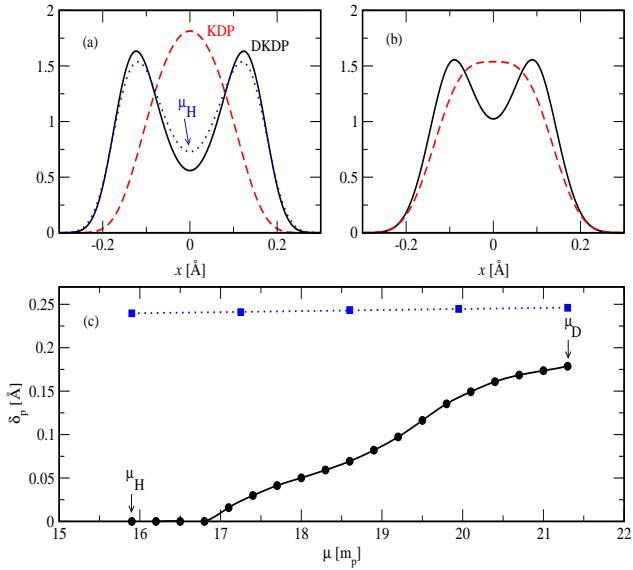},angle=270,width=9.cm,height=8.cm}}
\caption{WF in the 7-H(D) cluster PES for (a) {\it ab initio} and (b)
self-consistent model calculations. Solid (dashed) lines are for D (H).
Dotted line is for H in the DKDP PES. (c) WF peak separation $\delta_p$
as a function of the cluster effective mass $\mu$ (given in units of the
proton mass) for the self-consistent
model (circles) and for fixed DKDP potential (squares).
Lines are guides to the eye.}
\label{selfc}
\end{figure}

\section{PRESSURE EFFECTS}

Experiments under pressure carried out during the late eighties
showed, within error bars, that T$_c$ depends linearly on $\delta$
for different H-bonded ferroelectric materials.~\cite{Nel91}
Strictly speaking, the linear relation is verified for some of
these compounds within a restricted region of the T$_c$ vs.
$\delta$ plot. In the case of KDP, the linear behaviour appears to
extend up to a pressure of 17 Kbar, where T$_c$ vanishes. The need
for very high pressures ($\approx 60$ Kbar) to achieve a vanishing
T$_c$ for DKDP and other related H-bonded compounds, precluded the
generalization of the enunciated hypothesis, and motivated the
development of improved diamond anvil cells. Nevertheless, a
striking observation arises when the extrapolation of the linear
behaviour in the mentioned materials is carried down towards lower
values of T$_c$: the critical temperature appears to vanish for
all systems, deuterated and protonated, with one, two and
three-dimensional H-bond networks, around a seemingly universal
point where $\delta=\delta_c \approx 0.2$ \AA. ~\cite{Nel91}

In particular, the effect of deuteration can be reverted by
applying pressure, and the critical temperature of KDP can be
reproduced by compressing DKDP in such a way that the structural
parameter $\delta_{DKDP}$ assumes a value very close to that
measured in KDP at the initial pressure.~\cite{Nel91} This is
valid at all the measured pressures.

In this Section we explore the connection between pressure and
isotope effect by means of first-principles calculations and the
aid of the previously introduced self-consistent model for the
geometrical effect.~\cite{Col03} To this purpose, we first focus
on first-principles calculations, where we used the PW approach
explained in section II, combining ultra-soft (O and H) and
norm-conserving (K and P) pseudopotentials.~\cite{Van90,Tro91}
Using linear-response theory,~\cite {Bar01} the unstable
ferroelectric mode at the zone-centre was identified. This
corresponds mostly to H-atoms displacements with the pattern shown
in Fig. 3, with heavy ions displacing to a lesser extent. The
O-atoms are practically fixed in this mode.

The FE mode amplitude is identified with the H off-centering
coordinate ($x$). The total energy profile as a function of $x$
displays an effective double-well potential for the FE
mode.~\cite{giuseppe:phd} Thus, the potential can be characterized
by two parameters: the energy barrier E$_b$ between the stable and
the unstable ($x=0$) configurations, and the separation between
minima $\delta_m$. The values of E$_b$ vs. $\delta_m$ obtained
under different applied pressures are plotted in Fig. 8. A nearly
quadratic behaviour with simultaneous vanishing of E$_b$ and
$\delta_m$ is observed. Classically, ferroelectricity would
dissapear above $ \approx $ 100 Kbar. However, the critical
temperatures vanish at substantially lower critical pressures
P$_c$, which are isotope- dependent: 17 Kbar for KDP and 60 Kbar
for DKDP.~\cite{End99} This can be understood by considering the
quantum character of the nuclear dynamics. In fact, the zero-point
energy, which is larger for the proton, should lower the effective
energy barrier leading to lower critical pressures.

\begin{figure}
\vspace*{0.0 truecm}
\hspace*{0.3 truecm}
\centerline{
\psfig{figure={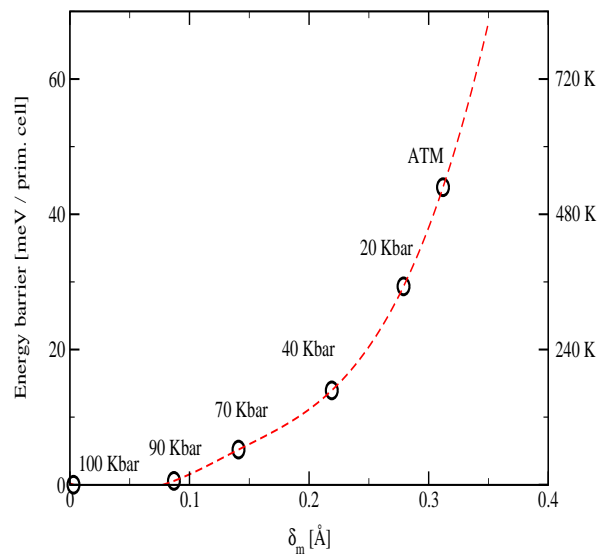},angle=270,width=9.cm,height=9.cm}}
\caption{Energy Barrier E$_b$ as a function of the distance between the
two minima of the double-well, $\delta_m$, for different pressures.
The dashed line is a  guide to the eye. An approximate conversion to
temperatures has been included to the right of the graph.}
\label{deltapress}
\end{figure}

In Fig. 9 we show the two parameters of the double-well, i.e.
E$_b$ and $\delta_m$, as a function of pressure. In this figure,
critical pressures for KDP and DKDP are indicated by vertical
arrows. These correspond, in our calculation, to different {\it
classical} values of the minima separation $\delta_m$, which are
around 0.3 \AA~ in KDP and 0.18 \AA~in DKDP. Experiments, however,
indicate that $\delta$ should approach the same universal value
$\delta_c \approx 0.2$ \AA~ for both compounds as T$_c$ goes to
zero. The difference found in the $\delta_m$ values should again
compensate for the quantum correction due to the nuclear dynamics.

\begin{figure}
\centerline{
\psfig{figure={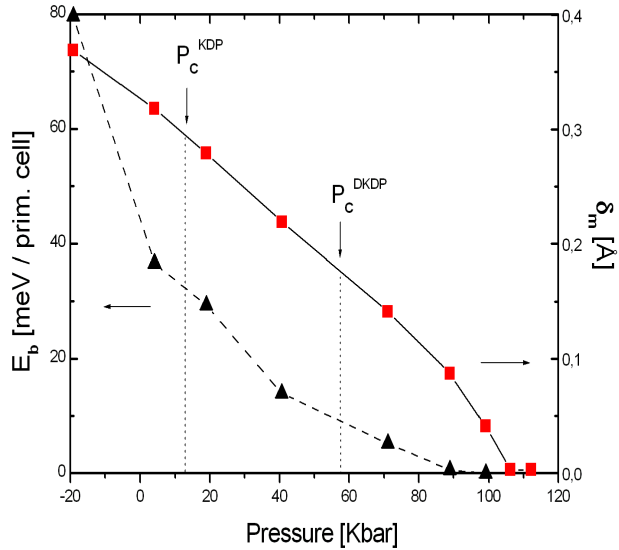},angle=270,width=9.5cm,height=8.5cm}}
\caption{Energy barrier E$_b$ (triangles) and double-well minima
separation $\delta_m$ (squares) as a function of pressure.
Lines are guide to the eye only.}
\label{figpress}
\end{figure}

In fact, being lighter, protons will localize closer to the middle
of the H-bond than deuteriums, leading in principle to such
compensation. However, we will show below that this effect alone
is not sufficient. The self-consistent geometrical effect
discussed in the previous section, which was essential to explain
the huge variation in the order parameter upon
deuteration~\cite{prlkov_02} is also crucial to explain the close
similarity of $\delta_c$ for deuterated and protonated systems.

Neutron diffraction experiments indicate that, to have the same
value of $\delta$, different pressures have to be applied to KDP
and DKDP.\cite{Nel91} When converting pressures into global energy
barriers using Fig. 9, the difference in energy barrier required
to have the same value of $\delta$ turns out to be nearly
independent of $\delta$, assuming a value of $ \approx 23 $ meV
per unit cell (two formula units). This energy difference then
seems to be a key quantity, which takes into account that the WF
of the more-easily-tunneling protons in KDP will exhibit the same
distance between peaks as the WF for deuterons in DKDP, only if
the underlying double-well is significantly deeper, i.e. by 23
meV, and also the distance between minima $\delta_m$ is increased.
The reason for this increased separation is in the very nature of
the hydrogen bond: the more distant the O-atoms, the less covalent
the bond, the larger $\delta_m$, and the deeper the double-well.
This is the essence of the geometrical effect, and 23 meV is the
energy difference required to modify the O-O distance in such a
way that the distance between peaks in the WF is the same for KDP
and DKDP.

To show how the geometrical effect enters into play, we considered
the feedback effective potential $V_{\rm eff}(x)$ from Eq.
\ref{efpot}. The probability distribution for the H(D) motion
$|\Psi(x)|^2$, was obtained by solving the Schr\"odinger equation
in the effective potential $V_{\rm eff}(x)$. We chose the N=7
cluster, with the relation $ \mu_D/\mu_H \approx 1.3 $ for the
effective masses in DKDP and KDP. The value of $\delta^{0}_m$ was
fixed to 0.24 \AA. Within this model, we studied how the value of
the energy barrier $E^{0}_b$ has to be modified (simulating the
application of pressure), in order to keep the peak separation
$\delta$ of the wave function $\Psi(x)$ constant upon deuteration.
This study was carried out for different values of the non- linear
parameter $k$ in the model. Large values of $k$ represent
important feedback geometric effects.

\begin{figure}
\vspace*{-1.0 truecm}
\centerline{
\psfig{figure={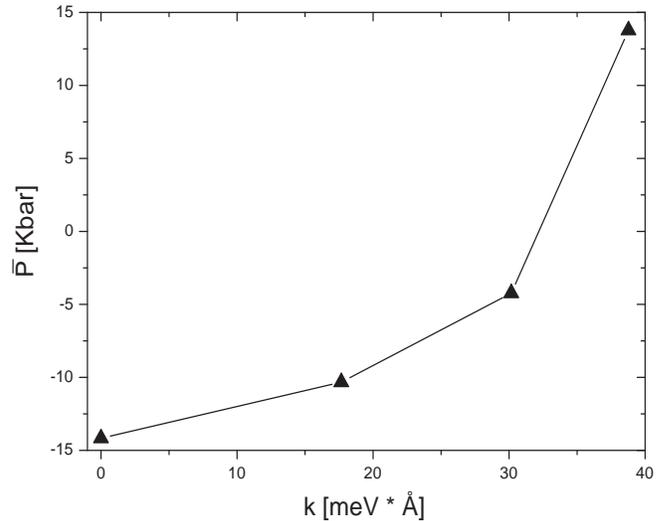},angle=0,width=10.cm,height=8.5cm}}
\caption{Average pressure $ \overline{P} $ as a function
of the coupling constant k from the non-linear model, as defined in text.
Lines are guide to the eye only.}
\label{figpk}
\end{figure}

For each value of the non-linear parameter $k$, we searched for
values of the barrier energies $E^{H}_b$ and $ E^{D}_b $ for each
isotope such that $\delta^{KDP} = \delta^{DKDP}$, with the
additional constraint that the difference $E^{H}_b-E^{D}_b = 23 $
meV remains constant. Using Fig. 9, we converted energy barriers
into pressure, with the warning that these correspond to global
energies, while $E^{0}_b$ represents cluster energy barriers.
Since here we are interested in qualitative issues, the former is
a reasonable approximation. Therefore, we calculate the average
pressure $\bar{P} = \left\lbrace P(E^{H}_b) + P(E^{D}_b)
\right\rbrace /2$ for each value of $k$, which serves to fix the
absolute scale of pressures necessary to fulfill the above
constraints. In Fig. 10 we plot $ \bar{P} $ as a function of $k$.
For a value of $k=0$ (no geometrical effect), in order to maintain
$\delta$ constant upon deuteration, we have to apply a large
negative pressure to the system, and hence expand it to have
significantly higher energy barriers. Conversely, for larger
pressures, as those experimentally measured, compensation can be
achieved only by considering large values of $k$, thus leading to
important structural non-linear effects. In summary, to explain
pressure effects on KDP and DKDP,\cite{Nel91} it is necessary to
consider the non-linear relation between isotope substitution and
geometric effect.

\section{DISCUSSION AND CONCLUSIONS}

The nature of the instability that drives the FE transition and
leads to the onset of the spontaneous polarization $ P_{s} $ has
been extensively discussed in the past.~\cite{Kob68,Bli87,popova}
Although the H(D) ordering in the basal plane is undoubtedly
correlated with the transition, it was originally assumed that $
P_{s} $, which is oriented along the c-axis, was due to the
displacements of K$^{+}$ and P$^{-}$ ions along this axis.
\cite{Kob68} However, the observed value of $ P_{s} $ can only be
explained within this model if unrealistic, very large charges for
the phosphorous ion are assumed. \cite{popova} Bystrov and
Popova~\cite{popova} proposed, alternatively, that the source of $
P_{s} $ could be the electron density shift in the P-O and P-O-H
bonds in the polar direction, which occurs when the protons order
perpendicularly. This assumption cannot be assessed through model
calculations, for it is originated in the complex electronic
interactions in the system.

By means of the present {\it ab initio} calculations we were able
to overcome this limitation, and to show that the FE instability
has its origin on an electronic charge reorganization within the
internal P-O and P-O-H bonds of the phosphates, as the H-atoms
order off-center in the H-bonds. As a matter of fact, the overall
effect produced by the H-ordering is an {\em electronic charge
flow} from the O2 side to the O1 side of the PO$_4$ tetrahedron,
and a concomitant distortion of the former. \cite{comp_mat_sci}
This is in agreement with the explanation given in Ref.
\cite{popova}, and also agrees with the results obtained by
another recent first-principles calculation. \cite{Zha02}

The microscopic origin of the global FE instability, i.e. the
connection between H ordering and phosphate distortions, is also
demonstrated in the strong correlation between the off-centering
parameters $ \delta $ for H and $ \gamma $ for K-P (see inset of
Fig. 4). One of the observations here is that the overall
potential for the H motion is not in fact separable, and one has
to deal with the problem of the "chicken and the egg", what is
really first? \cite{Kru90} Nevertheless, using that $ \gamma $ is
a measure of polarization, we showed that the source of the FE
instability is in the H off-centering, and not vice versa.

There is a long controversy about the origin of the FE transition.
Some experimental facts support the coupled proton-phonon model
which displays essentially a displacive-like transition.
\cite{Tok87} Other experiments, e.g. Raman studies \cite{Tom83},
seem to indicate the importance of the order-disorder character of
the transition originated in the H$_{2}$PO$_{4}$ unit dipoles.
Electron-nuclear double-resonance (ENDOR) measurements
~\cite{Rav91} indicate that not only the H$_{2}$PO$_{4}$ group,
but also the K atoms, are disordered over at least two
configurations in the paraelectric phase.  It is also shown by
neutron scattering experiments that the P atom is distributed over
at least two sites in DKDP.~\cite{McM90e,McM91} In spite of the
still unresolved character of the transition, it is clear that
local instabilities arising from the coupling of light and heavy
ions are very important in this system, irrespective of the
correlation length scale associated with the transition.

In fact, our calculations in the PE phase show that local proton
distortions with the FE mode pattern need to be accompanied by
heavy ion relaxations in the PO$_{4}$-K group to produce
significant instabilities, a fact which is in agreement with
experiments. We have shown that the correlation length associated
with the FE instability is much larger in KDP than in DKDP,
suggesting that DKDP will behave more as an order-disorder
ferroelectric than KDP.

The coherent interference of the proton in the two equivalent
sites of the PE phase was observed recently by neutron Compton
scattering experiments. ~\cite{Rei02} Quantum coherence arises in
our calculations for DKDP only when P and K ions are allowed to
relax together with the deuterons. In KDP, the onset of tunneling
and hence, coherence, would require relaxations of clusters
comprising more than three KH$_{2}$PO$_{4}$ groups, which were not
considered in the present work. The momentum distributions
calculated in the PE phase are in qualitative agreement with the
experiment.~\cite{Rei02} Quantum coherence would, thus, be
produced by a dressed proton, i.e. strongly correlated with the
heavier ions. In fact, in Ref.~\cite{Rei02}, many-body effects due
to the motion of the surrounding ions are not excluded, but it
turns out that these are difficult to assess. We have also found
good agreement with the experiment for the momentum distribution
in the FE phase, which corresponds to a single, anharmonic well
for each individual proton, in a host FE lattice. In this
situation there is no coherence between the motions of the various
protons.

The most striking feature, which is not yet satisfactorily
understood, is undoubtedly the huge isotope effect in the critical
temperature and the order parameter of the transition. The first
explanation was that proposed by the tunneling model and later
modifications,~\cite{Bli60,Kob68} but soon after the vast set of
experiments carried out by Nelmes and co-workers,
~\cite{Nel_rev_87,Nel88,McM90,McM90e,Nel91} and the comprehensive
structural compilation of Ichikawa {\it et al.},~\cite{Ich87} the
importance of the so- called geometrical effect as an alternative
explanation became apparent. Other
experiments~\cite{Tom83,Ike94,Rei02} and
models~\cite{Mat82,Sug91,Bus98} favoured one or the other vision,
or even both, but an overall and consistent explanation of the
phenomenon is still lacking. Still unanswered questions like: if
tunneling occurs, what are the main units that tunnel?, what is
the connection between tunneling and geometrical effects?, and
what is the true microscopic origin of the latter?, are possibly
some of the reasons why a full explanation of the isotope effect
is not yet available. With the aid of the {\it ab initio} scheme,
our efforts in this direction shed also light onto the underlying
microscopic mechanism for the isotope effect.

Protons alone are not able to tunnel because the effective
potentials in which they move display tiny double wells, and the
particles are broadly delocalized around the center of the
H-bonds. In this regard, we conclude that the simplified version
of the tunneling model, i.e. that of a tunneling proton, or even a
collective proton soft-mode alone, is not supported by our
calculations. On the other hand, we observe "tunneling clusters",
with an effective mass much larger than that of a single, or even
several, protons (deuterons), due to the correlation with heavier
ions. These clusters have different sizes, leading to different
lengths and energy scales competing in the system in the PE phase.
The smallest tunneling unit in DKDP is found to be the
KD$_{2}$PO$_{4}$ group. This results agree with the idea developed
by Blinc and Zeks of a tunneling model for the whole
H$_{2}$PO$_{4}$ unit,~\cite{Bli82} which helped to describe the
typical order-disorder phenomena observed in some experimental
trends. ~\cite{Tok87} However, the explanation for the isotope
effect arising in the present work is even more complex, and goes
beyond the concept of ``tunneling alone''as its main cause.

Although the PE phase of the system shows a complex scenario due
to the appearance of different length scales, it is clear that
larger clusters will prevail as the transition approaches. We have
shown for both isotopes that tunnel splittings in these clusters
at fixed potential are much smaller than the thermal energy at the
critical temperature. Thus, at fixed potential, tunneling is not
able to account for the large isotope effect in the system.
However, as the dressed particle is delocalized by tunneling, the
effective potential felt by the H(D)-atom changes upon isotopic
substitution, due to significant modifications in the chemical
properties of the O-H...O bond, which are reflected in a
concomitant {\it lattice relaxation}.

With the aid of a simple model based in our {\it ab initio}
results, we were able to show how this feedback effect strongly
amplifies the geometrical modifications in the H(D)-bridge.
Tunneling triggers a self-consistent mechanism, but in the end,
the geometrical effect dominates the scenario and accounts for the
huge isotope effect, in agreement with neutron scattering
experiments.~\cite{McM90,Nel91} Therefore, these aspects, which
were largely debated in the past, here appear as complementary and
deeply connected to each other.~\cite{Mat82,Bli87,prlkov_02}

The feedback effect of the geometrical modifications on the proton
distribution is also necessary to explain the results of
experiments under pressure. ~\cite{Nel91} There, it was observed
that the critical pressures at which the transition temperature
vanishes correspond to an isotope and material-independent value
of the peak separation in the proton distribution. We have shown
that this unique value for KDP and DKDP can be achieved only as a
consequence of a compensation between quantum delocalization
effects and geometrical modifications imposed by
pressure.~\cite{Col03}

The question of why T$_c$ is so closely related to the distance
between H peaks $\delta$ is still open. A possible explanation can
rely on the fact that equal GS levels relative to the top of the
barrier in a double well potential correspond to approximately
equal $\delta$, irrespective of the energy barrier height and mass
of the tunneling particle. We have verified in test calculations
that this holds as long as the GS energy is not very deep below
the top of the energy barrier, which means not too low pressures,
as discussed in Section VI. In addition, in our mean-field
description of the scenario of the FE transition, T$_c$ should be
related to the energy difference between the GS level and the top
of the barrier in the double well potential of a tunneling cluster
(see Fig. 5). This could explain the close relation between T$_c$
and $\delta$ observed in neutron diffraction
experiments.~\cite{Nel91}

The nonlinear feedback between tunneling and structural
modifications is a phenomenon of wider implications. Tunneling
units are indeed observed in a large variety of molecular
compounds and biomolecules. Both tunneling and structural changes
are important for the reaction mechanisms of enzymes~\cite{Koh99}
and other biological processes. Our results on KDP supports the
already expressed need for revision of the general theories of
host-and-tunneling systems.~\cite{Kru90}

In summary, we showed that proton ordering in KDP leads to an
electronic charge redistribution and ionic displacements that
originate the spontaneous polarization of the ferroelectric phase.
The instability process is controlled by the hydrogen
off-centering. The double-peaked proton distribution in the
bridges, observed in the paraelectric phase, cannot be explained
by a dynamics of protons alone. These must be correlated with
displacements of the heavier ions within clusters. These tunneling
clusters can explain the recent evidence of tunneling obtained
from Compton scattering measurements. We also showed that the mere
mass change upon deuteration does not explain the huge isotope
effect observed. We find that structural changes arising from the
modification of the covalency in the bridges produce a feedback
effect on the tunneling that strongly enhances the phenomenon. The
resulting influence of the geometric changes on the isotope effect
is in agreement with experimental data from neutron scattering.
Moreover, the behavior of the proton/deuteron distribution in the
bridges under pressure can only be explained by invoking the
mentioned feedback effect of geometry.

\section*{Acknowledgments.}

We thank E. Tosatti for helpful discussions. J. K. and G. C. thank C. S\'anchez for valuable discussions. S. K. also acknowledges very useful discussions with N. Dalal and R. Blinc. R. M. and S. K. thank support from CONICET and ANPCyT (grant PICT99 03-07248), Argentina and from ICTP, Trieste, Italy. S. K. also thanks support from Fundaci\'on Antorchas, Argentina and the Florida State University (FSU), Tallahassee, USA. G. C. thanks the European Social Fund for funding. Part of the calculations have been done using the allocation of computer time of the UKCP consortium.


\begin{references}
\bibitem{note} The minimum size leading to a tunneling state bellow
the central barrier is a measure of the crossover
length to a quantum dominated FE transition.
\end{references}
\end{document}